# High-Q aluminum nitride photonic crystal nanobeam cavities

W. H. P. Pernice[1*], C. Xiong[1], C. Schuck[1] and H. X. Tang[1†]

[1]*Department of Electrical Engineering, Yale University, New Haven, CT 06511, USA*

(February 7th, 2012)

We demonstrate high optical quality factors in aluminum nitride (AlN) photonic crystal nanobeam cavities. Suspended AlN photonic crystal nanobeams are fabricated in sputter-deposited AlN-on-insulator substrates using a self-protecting release process. Employing one-dimensional photonic crystal cavities coupled to integrated optical circuits we measure quality factors up to 146,000. By varying the waveguide-cavity coupling gap, extinction ratios in excess of 15 dB are obtained. Our results open the door for integrated photonic bandgap structures made from a low loss, wide-transparency, nonlinear optical material system.

[*] Current address: Institute of Nanotechnology, Karlsruhe Institute of Technology, 76133 Karlsruhe, Germany.

[†] Email: hong.tang@yale.edu

Over the last decade considerable research effort has been devoted to developing high quality factor (Q factor) photonic crystal (PhC) cavities that have dimensions comparable to the wavelength of light [1-4]. By shrinking the modal volume $V$ to near the fundamental limit of $V = (\lambda/2n)^3$, these cavities have enabled various applications ranging from ultrasmall lasers [5, 6], strong light-matter coupling [7-9] to optical switching [10]. Besides designs in two-dimensional PhC cavities, there has been much interest in cavities realized in suspended nanobeams patterned with a one-dimensional (1D) lattice of holes [11-14] . These 1D PhC cavities offer exceptional quality factor to modal volume ratios (Q/V), relative ease of design and fabrication, and show much potential for the exploitation of optomechanical effects [15, 16]. While such cavity designs have been investigated in stand-alone configurations in which the cavity is read out via free-space optical setups or fiber tapers, their true potential lies in the integratability with nanophotonic circuitry. While integrated photonic circuits have mostly been developed for applications in the telecoms window around 1550 nm, there are increasing interests in realizing nanophotonic circuits for applications in the near infrared and visible wavelength regime. Therefore, optical materials with wide transparency windows are investigated, such as silicon nitride (SiN), gallium phosphide (GaP), lithium niobate ($LiNbO_3$) and III-nitride semiconductors [17]. We previously showed that gallium nitride (GaN) thin-films, wafer-bonded to oxidized silicon wafers, provide a promising platform for nonlinear integrated optics [18]. However, the bonding process required for substrate preparation can lead to film-uniformity issues because the final surface has to be smoothed by chemical-mechanical polishing. Therefore, the obtainable quality factors in optical microresonators based on bonded GaN thin films are limited by the remaining surface roughness.

AlN, on the other hand, is a wide-bandgap semiconductor with the capability of direct integration on silicon (100) substrates. Additionally, AlN shows exceptional mechanical and thermal properties and has found wide use in microelectromechanical resonators due to its high piezoelectric transduction efficiency [19, 20]. Here we introduce AlN thin films on insulating silicon dioxide as a suitable optical material system for photonics applications. By using sputter deposition of c-axis oriented AlN, wafer-scale photonic circuitry can be fabricated with standard nanofabrication tools. By using a sputtering procedure, very low surface roughness can be obtained, leading to a superior substrate material compared to the aforementioned GaN structures. Furthermore, because the crystallographic orientation of the sputtered film is random in plane, direction dependent etching and resulting waveguide non-uniformieties can be avoided. We design and experimentally demonstrate high Q 1D photonic crystal cavities in free-standing AlN nanobeams by employing tapered Bragg mirrors. The cavity is integrated into an on-chip photonic circuit and coupled to a feeding waveguide for convenient optical access. We measure optical quality factors up to 146,000 and achieve high extinction ratio in critically coupled devices. Our AlN-on-Insulator (AOI) platform provides a viable route towards wideband optical applications in an integrated framework.

We investigate the cavity design illustrated schematically in Fig.1(a). In our layout the cavity region is enclosed between tapered PhC Bragg mirrors, in which the primary lattice constant of the PhC is tapered down parabolically towards a secondary, smaller lattice constant in the cavity region [13]. The cavity is patterned into a free-standing AlN (refractive index of 2.1) nanobeam with a height of 330 nm. Because AlN provides a lower refractive index contrast against air defined as $\Delta = (n_1^2 - n_2^2)/2n_1^2$ ($\Delta$=1.7) in comparison to high-index materials such as silicon

W.Pernice, C. Xiong, C. Schuck and H.Tang

($\Delta$=5.5), opening a robust photonic bandgap in-plane is hard to achieve. Therefore 1D PhC cavities provide superior quality factors than their two-dimensional counterparts. In this case lateral confinement is provided by the waveguide rather than the PhC as in 2D bandgap structures. Thus by exploiting Bragg reflection high quality cavities can be realized without the need for a high index contrast material system. We optimize the 1D PhC using the relevant geometric parameters as defined in Fig.1a), which are the length of the taper ($L_t$), the hole radii at the cavity center ($R_2$) and in the mirror region ($R_1$), as well as the waveguide width (*w*). The whole radii are linked to the lattice constant $l_i$ by keeping a fixed filling ratio, so that $R_i=f*l_i$. By scanning this multi-dimensional parameter space using a finite-difference time-domain (FDTD) method, we arrive at an optimized design space where the cavity parameters can be investigated experimentally. In the optimized design we use a primary lattice constant of $l_1$=560 nm which is tapered down to the cavity region parabolically over a length of 14 lattice periods to a value of $l_2$=430 nm. The ratio of the hole radius to the PhC lattice constant is kept at f=0.29, while the width of the nanobeam is set to 975 nm. By varying the distance between the inner-most holes (inner-hole spacing) in the optimized PhC design, intrinsic optical quality factors up to $1.1\times10^6$ are found numerically. Changing the inner-hole spacing, however, also affects the cavity resonance condition, thus allowing for tuning of the resonance wavelength. When the feeding waveguide is included in the simulations, the resulting optical quality factors are reduced depending on the external coupling strength, determined by the gap between the waveguide and the cavity beam. Under critical coupling conditions, the cavity linewidth increases by a factor of roughly 2 and thus the optical Q amounts to half of the intrinsic value.

W.Pernice, C. Xiong, C. Schuck and H.Tang

To confirm the predicted optical properties of the designed PhC structures, nanophotonic devices are fabricated from AlN-on-insulator substrates. A 330 nm thick AlN film is deposited onto silicon wafers with a 2.6 µm thick buried oxide layer using a sputtering process. The sputter deposition leads to c-Axis oriented, polycrystalline AlN thin films with low residual stress. The film thickness of both the waveguiding layer and the buried oxide layer was chosen such that maximal coupling efficiency of the grating couplers into the feeding waveguide is achieved. In the optimized design constructive interference between incoming light and light reflected from the underlying silicon carrier wafer leads to a maximum coupling efficiency of roughly 30 % for each coupler. We use a two-step electron-beam (e-beam) lithography process with subsequent reactive ion etching (RIE) to define the cavity and the supporting on-chip photonic circuitry, as illustrated in Fig.1(c). A first iteration of e-beam lithography employing hydrogen silsesquioxane (HSQ) e-beam resist is used to mask the designed optical circuit against the subsequent etch step (i)). A timed RIE in $Cl_2/BCl_3/Ar$ inductively coupled plasma is applied to etch the patterns into the AlN film, leaving a residual AlN slab of 70 nm (ii)). Because the AlN thin film does not have a crystallographic preference in plane direction-dependent dry etching is not observed in the fabricated structures. In a second e-beam lithography step using the positive resist ZEP520A, we define release windows over the cavity region (iii)). Then a second RIE is performed to remove the residual AlN slab in the cavity region (iv)) and also to etch all the way through the PhC holes. During the etch step the cavity beam is still protected by the residual HSQ resist which is left over from the first etch. The final waveguide profile shows a sloped sidewall angle of less than 8 degrees, extracted from optical inspection of scanning electron micrograph (SEM) images. After removal of the ZEP resist, the 70 nm AlN slab remaining everywhere except for the cavity region provides a natural mask layer against subsequent wet etching. A timed wet etch step in buffered oxide etchant is performed in order to remove the underlying buried oxide and release the nanobeam cavity (v)). The wet etching leads to a sufficient undercut of the cavity beam such that the influence of the substrate is minimized during optical measurements. An optical micrograph of a fabricated sample is shown in Fig.1(c), with nanophotonic waveguides as well as alignment markers which indicate the cavity region (vertical bars in the top half of the image). The grating couplers are used to launch light into the feeding waveguide, which is routed to the cavity region as shown in the SEM picture in Fig.1(d). A zoom into the cavity region, see Fig.1(e), illustrates that the devices can be fabricated with low sidewall roughness and almost vertical angles. In order to assess the remaining sidewall roughness we perform atomic force microscopy along the sides of the waveguides. From the measured profile we extract a residual surface roughness of 3.5 nm after etching. Compared to the operating wavelength around 1550 nm the roughness value does not provide significant scattering centers and thus allows for the fabrication of high quality photonic structures. The width of the in-coupling nanophotonic waveguides is kept at 1000 nm in order to allow for single-mode operation. The width of the cavity beam on the other hand is fixed at 975 nm, which yielded the highest Q values in the numerical simulations.

The optical properties of the fabricated device are investigated by measuring the transmission through the feeding waveguide. Light from a tunable laser source (New Focus 6428) is coupled into the chip using an optical fiber array. The chip is mounted on a motorized stage in order to allow for efficient alignment of the fibers to the grating couplers. After transmission through the chip, the light is collected with a low-noise photodetector (New Focus 2011). In order to optimize the coupling to the cavity we fabricate a variety of structures with varying gap between the feeding waveguide and the cavity beam. In the final design, the feeding waveguide is an arc with a radius of 10 μm in order to avoid significant bend loss. Because the optical mode profile of the nanobeam is highly concentrated in the cavity region as shown in Fig.2(a), the feeding waveguide is designed to only couple to the nanobeam near the point of highest intensity. By measuring the transmission profile of the feeding waveguide the cavity properties can be extracted. A typical result for a device with a coupling gap of 350 nm and an inner-hole spacing of 600 nm is shown in Fig.2(b) The envelope of the grating coupler's transmission profile exhibits small fringes due to back-reflection from the output grating coupler. A clear dip at 1530.9 nm in the transmission spectrum indicates the cavity resonance wavelength. Fitting the resonance with a Lorentzian function as shown in the detailed dataset of Fig.2(b) reveals a loaded cavity Q of 85,000.

In order to evaluate the influence of the feeding waveguide we fabricate several devices with coupling gaps ranging from 200 nm to 700 nm. The measurement results are shown in Fig.2(c). For the smallest coupling gap we find an overloaded cavity with an expected high extinction ratio greater than 15 dB. By increasing the separation between input waveguide and nanobeam the coupling strength reduces exponentially, which goes in hand with a significant reduction of the cavity linewidth. In this case the cavity is operating in the undercoupled regime. The reduction in coupling strength is accompanied by a cavity red-shift as shown in Fig.2(c). The red-shift is expected due to the decrease of the effective refractive index of the cavity mode when the input beam is moved further away from the cavity region.

In order to assess the photonic performance of the resonators we measure the surface roughness of the waveguides as well as the spectral properties. An atomic force microscope (AFM) image taken across a nanophotonic waveguide reveals a residual root-mean-square (rms) roughness (see Fig.3(a)) of the top surface of 1.2 nm. The AFM image shows a rounded waveguide profile due to the radius of the AFM tip, however, the sidewall roughness is still reproduced by the scanning procedure. The measured surface rms-roughness is smaller than the sidewall roughness (3.5 nm) as a result of the RIE etching procedure, confirming that the sputtering process leads to a very smooth substrate material. To investigate the spectral dependence of the cavity mode on the geometry of the cavity we then measure a series of devices with varying inner-hole spacing, around the designed optimum gap of 600 nm, as shown in Fig.3(b). The experimentally determined Q is indicated by the blue markers, while the numerical results are shown by the dashed lines. The best optical Q is reached for an inner-hole spacing of 600 nm, where we measure optical Q of 146,000. A zoom into the resonance for the highest Q device is shown in Fig.3(c), revealing a cavity linewidth of 10.5 pm for a resonance wavelength of 1530.62 nm. When the inner-hole spacing is reduced or increased, the optical Q drops significantly. The trend is also confirmed by numerical simulations of the intrinsic cavity design, as shown by the dashed blue line in Fig.3(b). When the feeding waveguide is added to the simulation we expect a critically coupled cavity Q of 340,000 from the FDTD simulations for the optimal cavity length of 600 nm. In the critical coupled case the intrinsic Q is reduced because the input waveguide provides an additional loss channel to the cavity. The lower quality factor found experimentally is furthermore attributed to the presence of the substrate in the evanescent cavity field and the surface roughness due to fabrication imperfections. We would like to point out, however, that the cavity geometry with highest optical quality factor can also be shifted to

W.Pernice, C. Xiong, C. Schuck and H.Tang

different wavelengths by scaling the optimized design accordingly. Fig.3(d) illustrates how the resonance wavelength increases with increasing inner-hole spacing. For the smallest length of 500 nm we find an optical resonance at 1523 nm while the resonance wavelength shifts to 1538.7 nm at a cavity length of 700 nm, i.e. by more than 15 nm. The experimentally found results are in good agreement with the numerical predictions, which are shown in Fig.3(d) by the dashed red line.

In conclusion we have realized integrated optical circuits in aluminum nitride thin films and demonstrated high-Q photonic crystal cavities in free-standing AlN nanobeams with a measured Q factor of 146,000, exceeding previously demonstrated quality factors in two-dimensional structures by far. By using sputter deposited AlN thin films, advanced substrate preparation techniques such as epitaxial growth or bonding processes can be avoided, thus making our AOI platform a promising candidate for integrated optics. Due to AlN's excellent linear and nonlinear optical properties, the results presented here could facilitate the development of on-chip frequency conversion components. Owing to the piezoelectric properties of AlN, our results could also lead to new optomechanical and tunable optical devices

This work was supported by a seedling program from DARPA/MTO and the DARPA/MTO ORCHID program through a grant from AFOSR. W.H.P. Pernice acknowledges support by the DFG grant PE 1832/1-1. H.X.T acknowledges support from a Packard Fellowship in Science and Engineering and a CAREER award from the National Science Foundation. We want to thank Dr. Michael Rooks and Michael Power for their assistance in device fabrication.

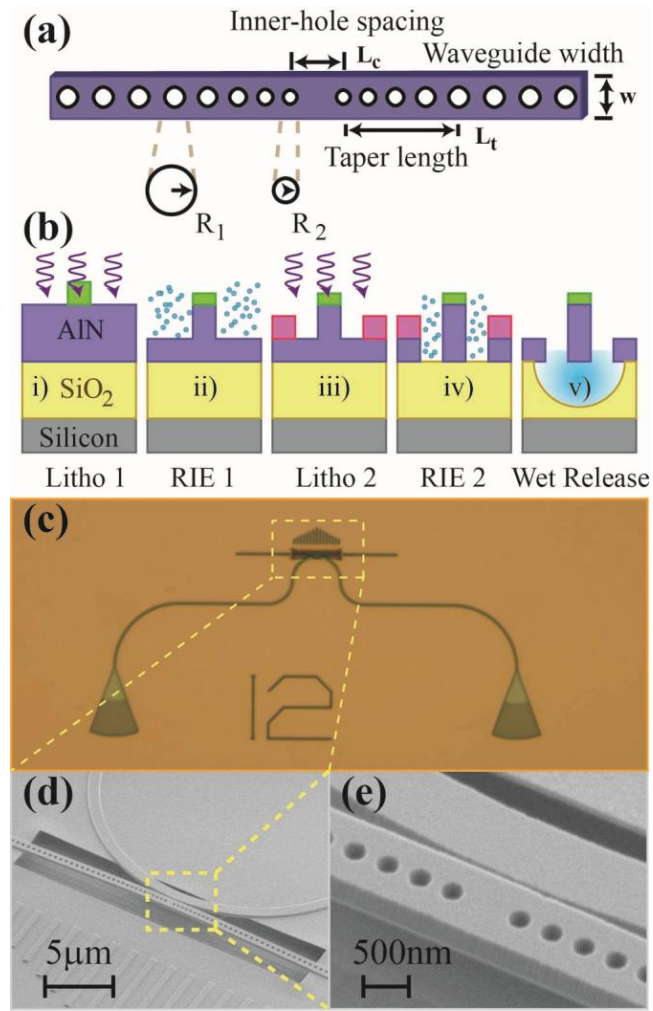

W.Pernice, C. Xiong, C. Schuck and H.Tang

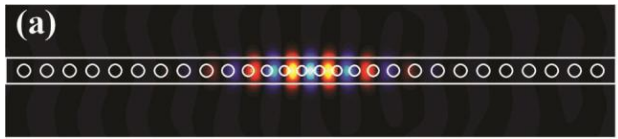
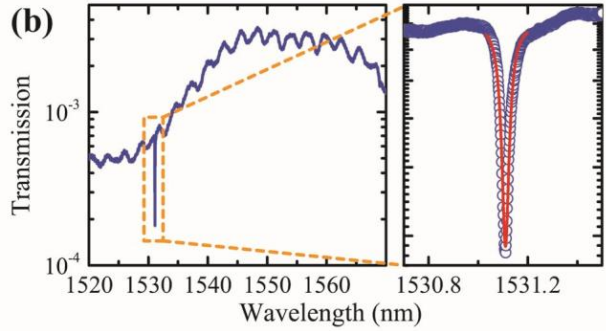
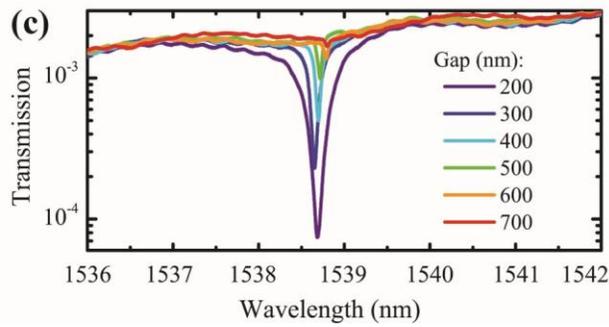
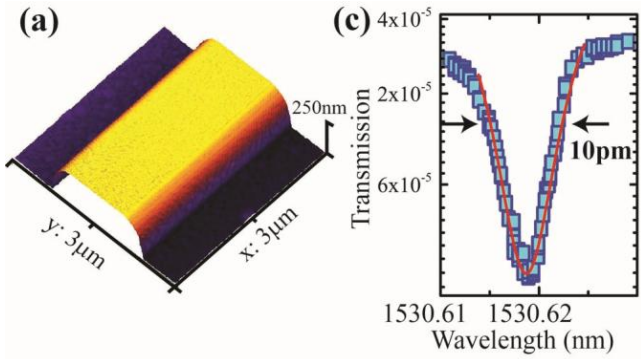
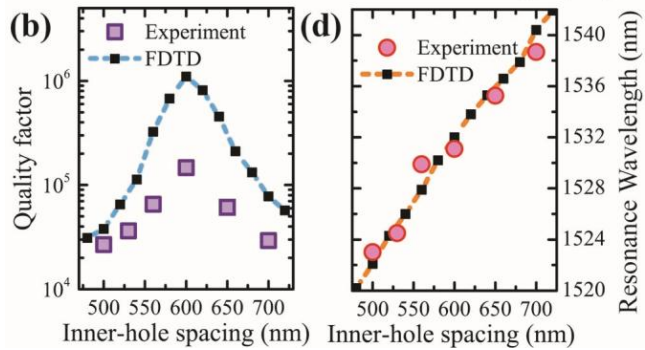

Fig. 1. a) The design of the one-dimensional AlN photonic crystal cavity. The hole radius is tapered parabolically from $R_1$ to $R_2$. b) The process flow used to fabricate the PhC devices. c) An optical micrograph of a fabricated photonic circuit with grating coupler input ports, nanophotonic waveguide and released cavity region. d) A SEM picture of the released waveguide region, showing the PhC nanobeam as well as the input waveguide. e) A magnified view of the cavity section of the waveguides.

Fig.2. (a) The simulated optical mode profile for the fundamental cavity mode. Shown is the x-component of the electric field. b) The measured transmission spectrum of a fabricated device, showing the cavity resonance at 1530 nm. The Lorentzian fit to the data (red line) in the zoom-in picture on the right reveals an optical Q of 85,000. c) The measured optical linewidth of the cavity in dependence of the coupling gap to the feeding waveguide.

Fig.3. (a) AFM image of the waveguide profile showing top surface rms-roughness of 1.2nm after fabrication and 3.5 nm sidewall roughness. (b) Tuning of the cavity resonance in dependence of the inner-hole spacing. The best optical Q of 146,000 is found for an inner-hole spacing of 600 nm. (c) A detailed spectrum of the cavity resonance with highest Q-factor of 146,000. (d) The resonance wavelength increases with inner-hole spacing. A length change from 500 nm to 700 nm shifts the resonance wavelength by 15 nm.